\documentclass{mem}
\usepackage{natbib}\usepackage{txfonts}\usepackage{balance}
\usepackage{graphicx,epsfig}

\begin{document}
\def\chandra{{\it Chandra}}
\def\xmm{{\it XMM-Newton}}
\def\asca{{\it ASCA}}
\def\sax{{\it BeppoSAX}}
\def\wfxt{{\it WFXT}}
\def\swift{{\it SWIFT}}
\def\xenia{{\it XENIA}}
\def\rosat{{\it ROSAT}}
\def\suzaku{{\it Suzaku}}

\title{
X-ray observations of cluster outskirts: \\
current status and future prospects.}
\subtitle{}

\author{
S. \,Ettori\inst{1,2}
\and S. \, Molendi\inst{3}
}

\offprints{S. Ettori, S. Molendi}

\institute{
Istituto Nazionale di Astrofisica --
Osservatorio Astronomico di Bologna, Via Ranzani 1, I-40127 Bologna, Italy
\email{stefano.ettori@oabo.inaf.it}
\and INFN, Sezione di Bologna, viale Berti Pichat 6/2, I-40127 Bologna, Italy
\and INAF, IASF, via Bassini 15, I-20133 Milano, Italy
\email{silvano@iasf-milano.inaf.it}
}

\authorrunning{Ettori \& Molendi}

\titlerunning{The outer regions of X-ray galaxy clusters}

\abstract{

Past and current X-ray mission allow us to observe only a fraction of the volume
occupied by the ICM. After reviewing the state of the art of cluster outskirts observations
we discuss some important constraints that should be met when designing an experiment 
to measure X-ray emission out to the virial radius. From what we can surmise \wfxt ~
is already designed to meet most of the requirements and should have
no major difficulty in accommodating the remaining few.

}
\maketitle{}

\section{Introduction}

Galaxy clusters form through the hierarchical accretion of cosmic matter.
The end products of this process are virialized structures that
feature, in the X-ray band, similar radial profiles of the surface brightness
$S_{\rm b}$
 (e.g. Vikhlinin et al. 1999, Neumann 2005, Ettori \& Balestra 2009)
and of the plasma temperature $T_{\rm gas}$
(e.g. Allen et al. 2001, Vikhlinin et al. 2005, Leccardi \& Molendi 2008).
Such measurements have definitely improved in recent years thanks to
the arcsec resolution and large collecting area of the present X-ray satellites,
like \chandra\ and \xmm, but still remain difficult, in particular
in the outskirts, because of the low surface brightness associated to these regions.
Present observations provide routinely reasonable estimates of the gas density, $n_{\rm gas}$, and
temperature, $T_{\rm gas}$, up to about $R_{2500}$ ($\approx 0.3 R_{200}$;
$R_{\Delta}$ is defined as the radius of the sphere that encloses a mean mass density
of $\Delta$ times the critical density at the cluster's redshift;
$R_{200}$ defines approximately the virialized region in galaxy clusters).
Only few cases provide meaningful
measurements at $R_{500}$ ($\approx 0.7 R_{200}$) and beyond
(e.g. Vikhlinin et al. 2005, Leccardi \& Molendi 2008,
Neumann 2005, Ettori \& Balestra 2009).
Consequently, more than two-thirds of the typical cluster volume,
just where primordial gas is accreting and dark matter halo is
forming, is still unknown for what concerns both its mass distribution
and its thermodynamical properties.
This poses a significant limitation in our ability to characterize 
the physical processes presiding over the formation and evolution of clusters 
and to use clusters as cosmological tools, as also outlined in the Scientific Justification
for the \wfxt\ (Giacconi et al. 2009).
Indeed the characterization of thermodynamic properties at large radii
would allow us to provide constraints on the virialization process, while 
measures of the metal abundance would allow us to  gain insight on the 
enrichment processes occurring in clusters (e.g. Fabjan et al. 2010).
Morever the X-ray emission at large radii could also be used to improve significantly 
measures of the gas and total gravitating masses thereby opening the way
to a more accurate use of galaxy clusters as cosmological probes (e.g. Voit 2005).

In these proceedings, we take stock of the situation on cluster outskirts
and suggest how to make progress.
In Sect.~\ref{sec: obs},  we provide an observational overview of 
currently available measures of cluster outer regions,  while in Sect.~\ref{sec: map}
we discuss some important constraints that should be met  when designing an experiment 
to measure X-ray emission out to the virial radius. 
In Sect.~\ref{sec: future}, we present an overview of future missions 
which have cluster outskirts observations as one of their 
goals, our main results are recapitulated  in Sect.~\ref{sec: summary}.

\begin{figure*}
\begin{center}
 \includegraphics[height=4.5cm]{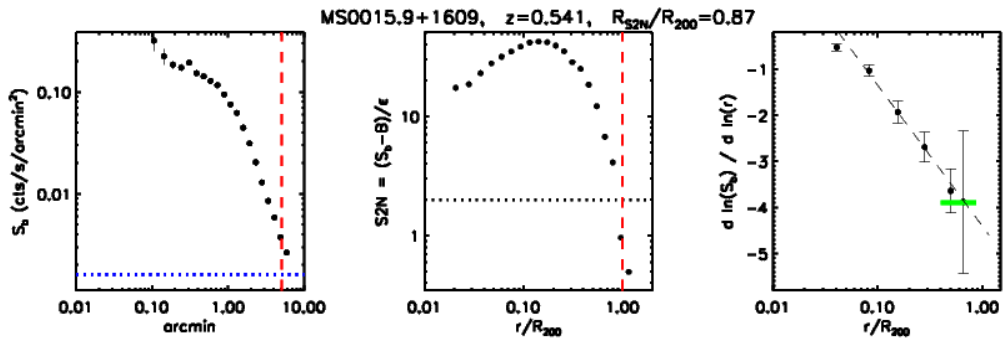}
\end{center}
\caption{\small
{\bf From left to right}: Example of a surface brightness profile
with the fitted background ({\it horizontal dotted line})
and the radius $R_{200}$ ({\it vertical dashed line});
the signal-to-noise profile evaluated as $S2N = (S_b - B)/ \epsilon$,
where the error $\epsilon$ is the sum in quadrature of the Poissonian error
in the radial counts and the uncertainties in the fitted background, $B$;
the best-fit values of the slope of the
surface brightness profile as a function of $r / R_{200}$. These
values are estimated over 6 radial bins
(thick horizontal solid line: the slope evaluated between
$0.4 \times R_{200}$ and $R_{S2N}$ with a minimum of 3 radial bins;
dashed line: best-fit of $d \ln (S_b) / d \ln (r/R_{200})$
with the functional form $s_0 +s_1 \ln (r/R_{200})$ over the radial range
$0.1 \times R_{200} - R_{S2N}$, with the best-fit
parameters quoted in Table~3 of Ettori \& Balestra 2009).
} \label{fig:s2n}
\end{figure*}

A Hubble constant of 70 $h_{70}$ km s$^{-1}$ Mpc$^{-1}$ in a flat universe
with $\Omega_{\rm m}$ equals to 0.3 is assumed throughout this manuscript.

\section{What we know of  cluster outskirts}\label{sec: obs}

\begin{figure*}
\begin{center}
 \includegraphics[height=5cm]{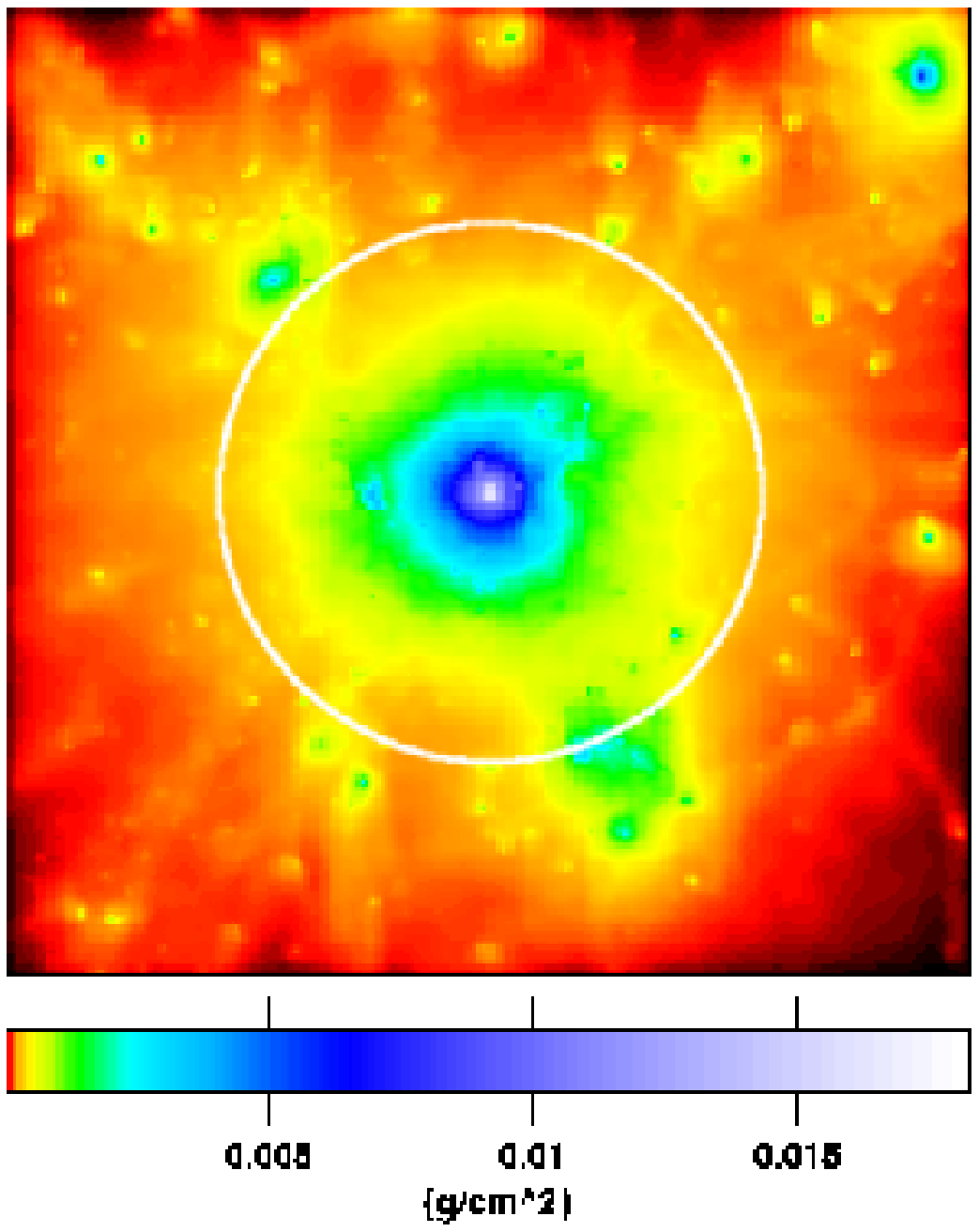}
 \includegraphics[height=5cm]{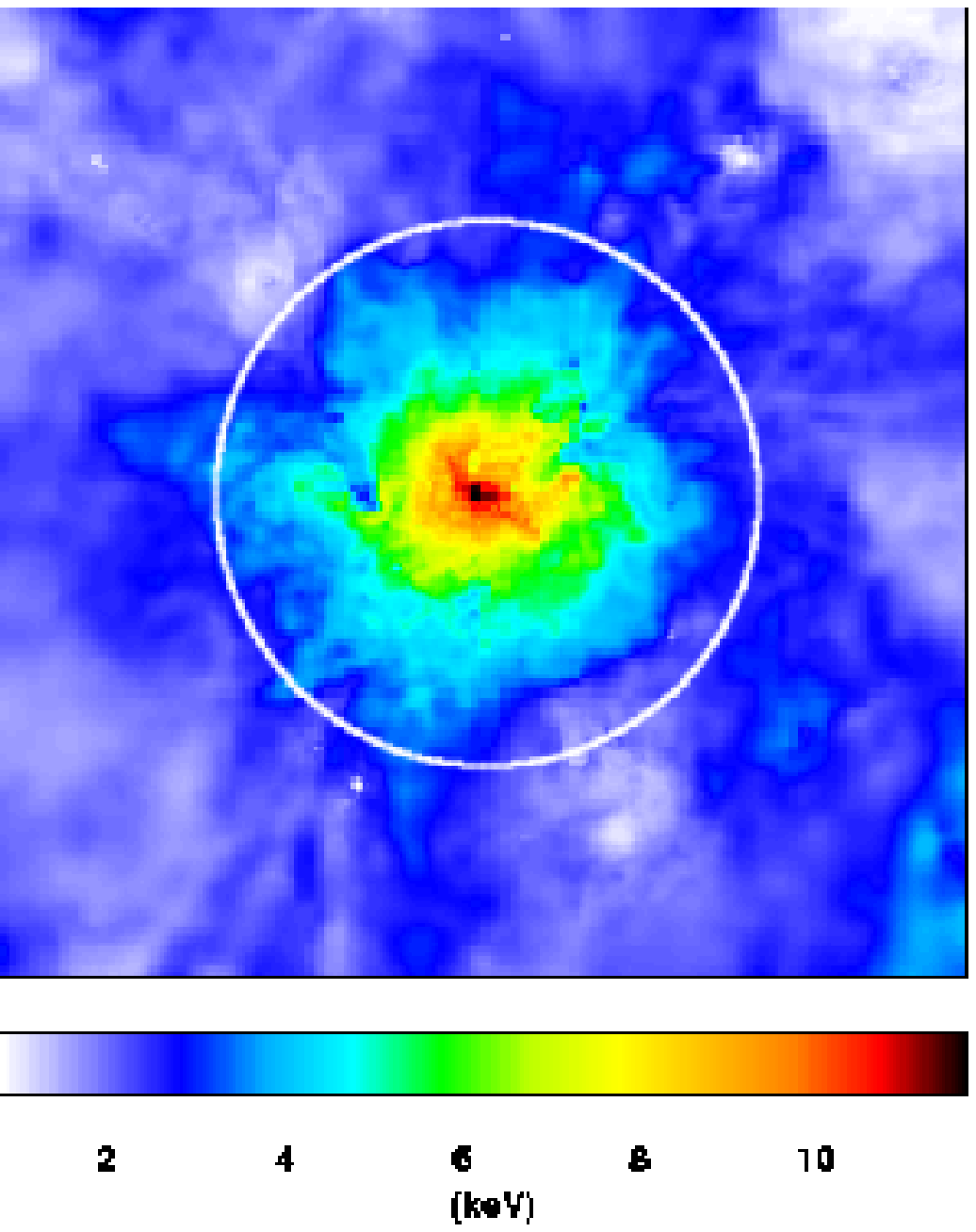}
 \includegraphics[height=5cm]{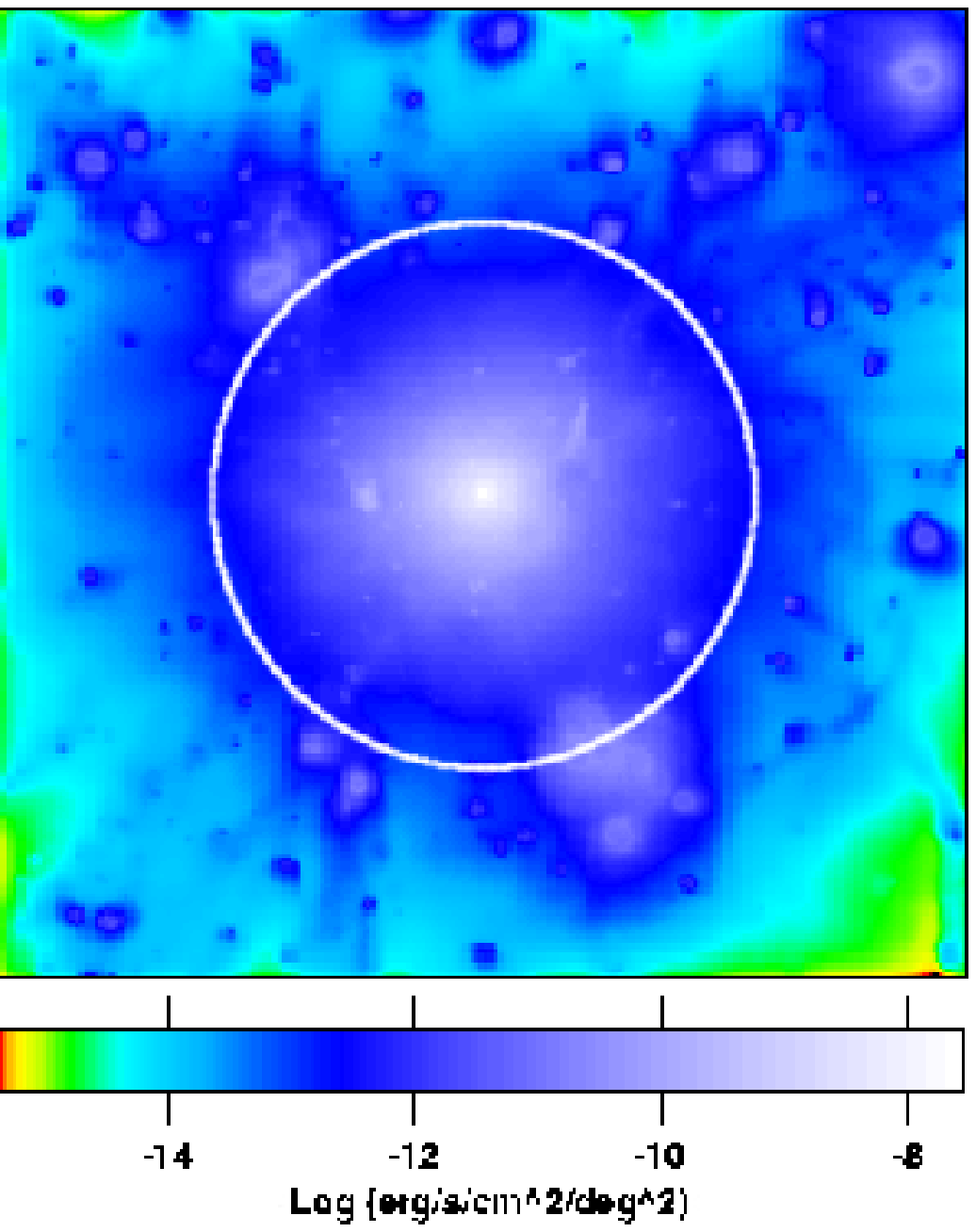}
\end{center}
\begin{center}
 \includegraphics[height=3.17cm]{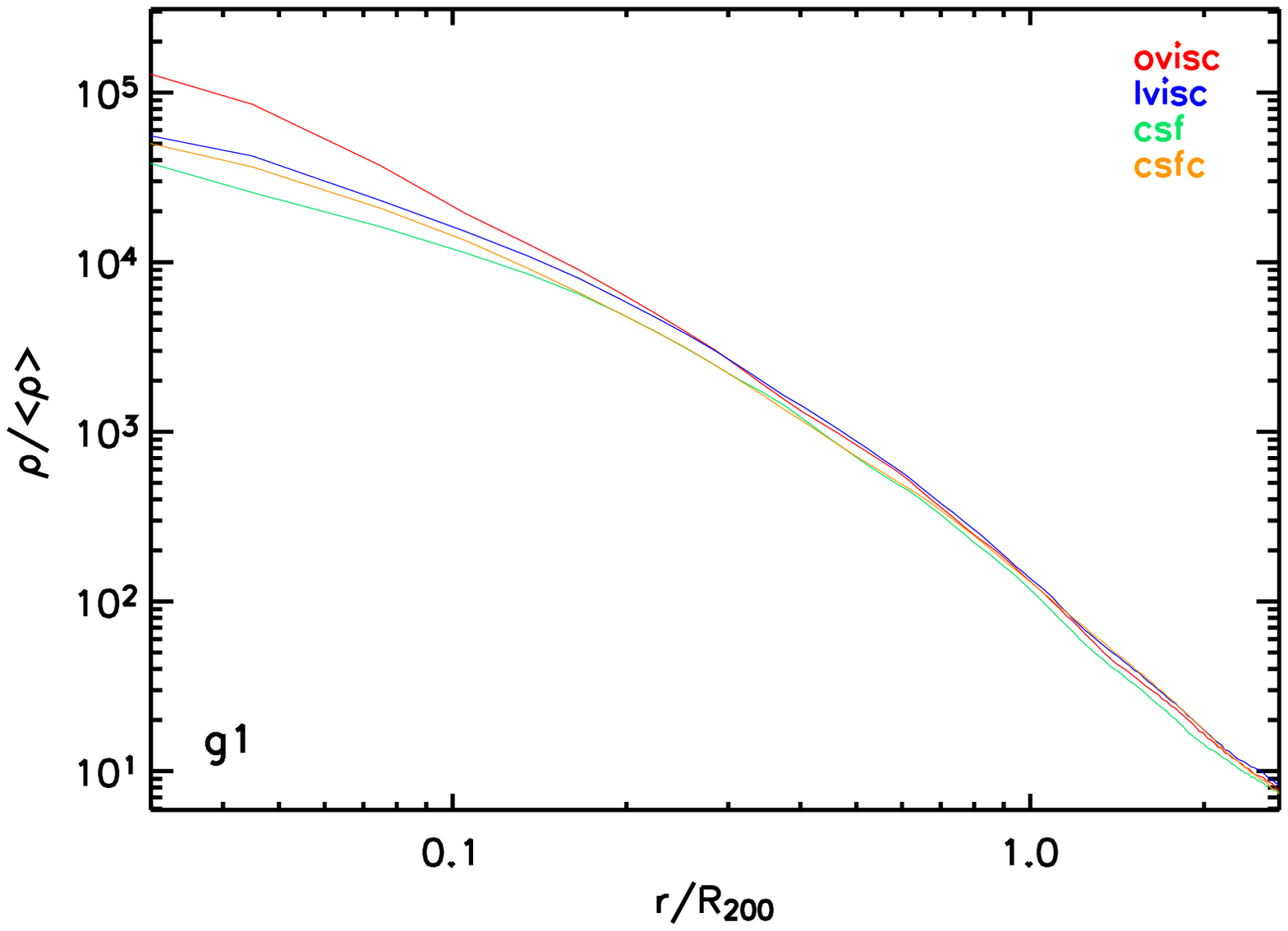}
 \includegraphics[height=3.17cm]{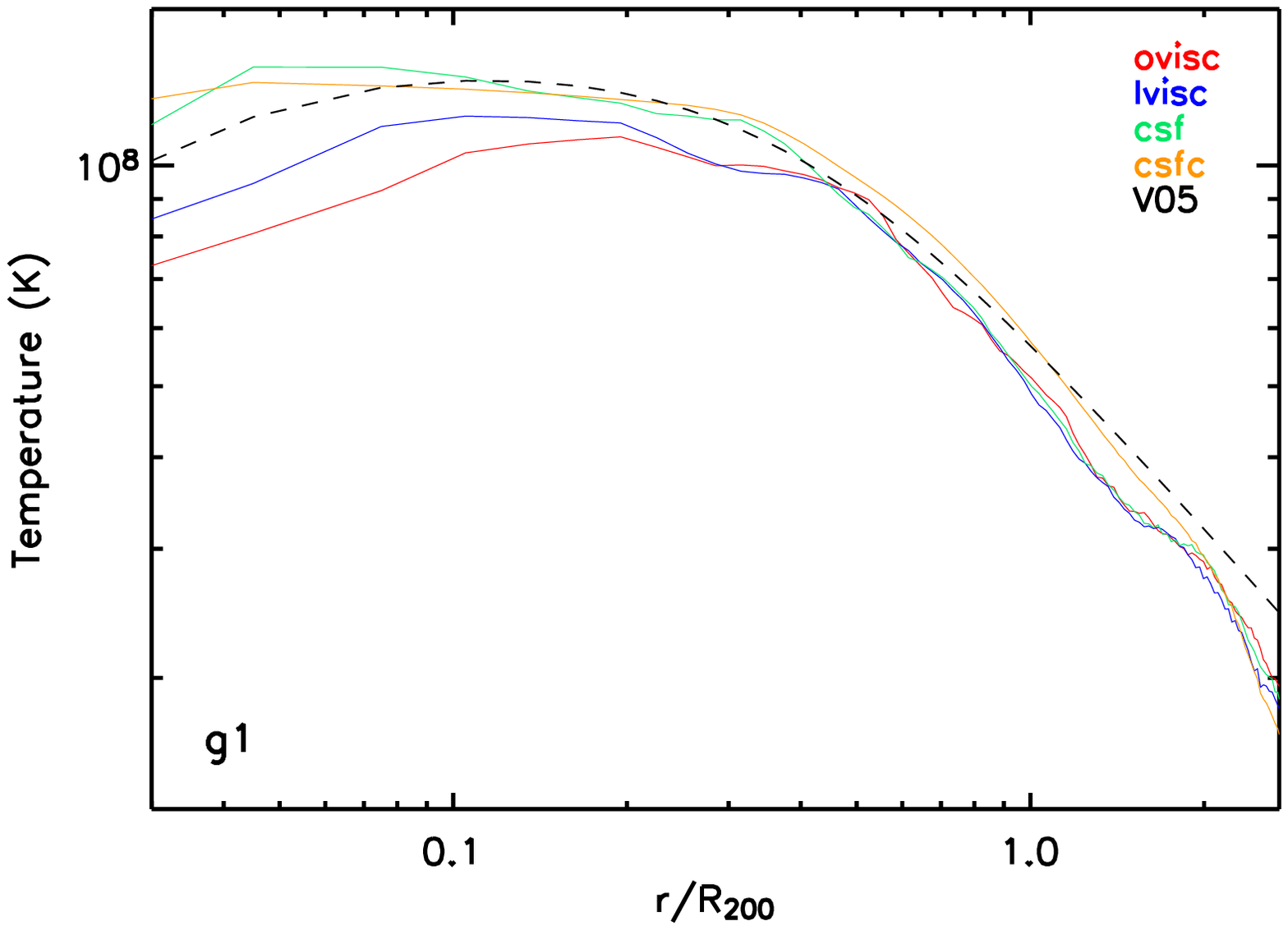}
 \includegraphics[height=3.17cm]{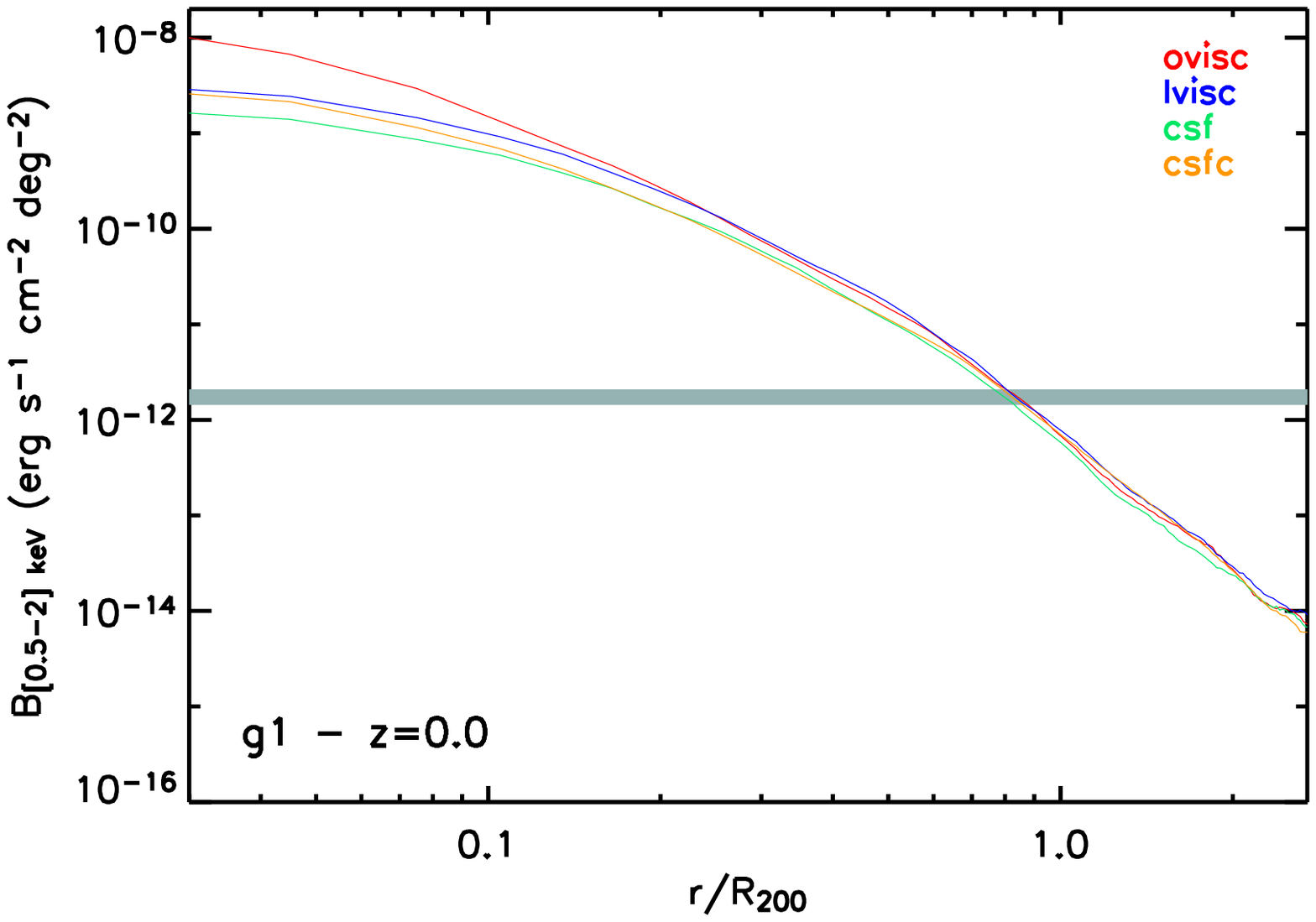}
\end{center}
\caption{\small  {\bf From left to right; upper panels}:
Maps of the projected gas density, mass-weighted temperature and soft
(0.5-2 keV) X-ray emission. The circles indicate the virial radius.
The size of the side of each map is 12 Mpc,
so they cover roughly up to $2.5 R_{200}$.
{\bf Bottom panels}:
Comparison between the gas density, mass-weighted temperature, soft
X-ray surface brightness profiles for a cluster with $M_{\rm vir}~2\times10^{15} M_{\odot}$
simulated by using 4 different physical models (Roncarelli et al. 2006). 
A dashed line indicates the functional
from Vikhlinin et al. (2006, eq. 9) that well reproduces the behavior
of the temperature profile of nearby bright galaxy clusters observed with \chandra.
The extragalactic unresolved background from Hickox \& Markevitch (2006) in the soft X-ray band
is indicated by the shaded region.
} \label{fig:simu}
\end{figure*}

\subsection{Surface brightness and gas density profiles}\label{sec: sb}

The X-ray surface brightness is a quantity much easier to characterize than
the temperature and it is still rich in physical information being proportional
to the emission measure, i.e. to the gas density, of the emitting source.
Recent work focused on a few local bright objects for which
{\it ROSAT} PSPC observations with low cosmic background and
large field of view have allowed to recover the X-ray
surface brightness profile over a significant fraction of the virial radius
(Vikhlinin et al. 1999, Neumann 2005).

In Ettori \& Balestra (2009), 
we study the surface brightness profiles extracted from 
a sample of hot ($T_{\rm gas} > 3$ keV), high-redshift ($0.3 < z < 1.3$) 
galaxy clusters observed with \chandra\ and described in Balestra et al. (2007).
A local background, $B$, was defined for each exposure by considering
a region far from the X-ray center that covered a significant portion
of the exposed CCD with negligible cluster emission.
We define the ``signal-to-noise'' ratio, $S2N$, to be the ratio of the observed surface
brightness value in each radial bin, $S_b(r)$, after subtraction of the estimated background,
$B$, to the Poissonian error in the evaluated surface brightness,
$\epsilon_b(r)$, summed in quadrature with the error in the background, $\epsilon_B$:
$S2N(r) = \left[ S_b(r) - B \right] / \sqrt{\epsilon_b(r)^2 +\epsilon_B^2}$.
The outer radius at which the signal-to-noise ratio remained above $2$ was defined
to be the limit of the extension of the detectable X-ray emission, $R_{S2N}$.
We estimated $R_{200}$ using both a $\beta-$model that reproduces
the surface brightness profiles and the scaling relation quoted in
eq.~\ref{eq:r200} and selected the 11 objects with $R_{S2N}/R_{200} > 0.7$
to investigate the X-ray surface-brightness profiles
of massive clusters at $r > R_{500} \approx 0.7 R_{200}$.
Examples of the analyzed dataset are shown in Fig.~\ref{fig:s2n}.
We performed a linear least-squares fit between the logarithmic values of the
radial bins and the background-subtracted X-ray surface brightness.
Overall, the error-weighted mean slope is $-2.91$ (with a standard deviation
in the distribution of $0.46$) at $r > 0.2 R_{200}$ and $-3.59 (0.75)$
at $r > 0.4 R_{200}$.
For the only 3 objects for which a fit between $0.5 R_{200}$ and $R_{S2N}$
was possible, we measured a further steepening of the profiles, with a
mean slope of $-4.43$ and a standard deviation of $0.83$.
We also fitted linearly the derivative of the logarithm $S_b(r)$
over the radial range $0.1 R_{200} - R_{S2N}$, excluding in this way the
influence of the core emission.
The average (and standard deviation $\sigma$) values of the extrapolated slopes are
then $-3.15 (0.46)$, $-3.86 (0.70)$, and $-4.31 (0.87)$
at $0.4 R_{200}$, $0.7 R_{200}$ and $R_{200}$, respectively.

These values are comparable to what has been obtained in recent analyses.
Vikhlinin et al. (1999) find that a $\beta-$model with $\beta=0.65-0.85$
describes the surface brightness profiles in the range $0.3-1 R_{180}$ of
39 massive local galaxy clusters observed with {\it ROSAT} PSPC.
For a $\beta-$model with $x=r/r_{\rm c}$, $\partial \ln S_b / \partial \ln x =
(1 - 6 \beta) \, x^2 / (1+x^2)$ and $\partial \ln n_{\rm gas}
/ \partial \ln x = -3 \beta \, x^2 / (1+x^2)$, impling that
$\beta=0.65-0.85$ corresponds to a logarithmic slope of the
surface brightness of $-2.9/-4.1$, that is a range that includes 
our estimates.
Neumann (2005) finds that the stacked profiles of few massive nearby systems
located in regions at low ($<6 \times 10^{20}$ cm$^{-2}$) Galactic absorption
observed with {\it ROSAT} PSPC still provide values of $\beta$ around $0.8$ at
$R_{200}$, with a power-law slope that increases from $-3$ when the fit is
performed over the radial range $[0.1, 1] R_{200}$ to $-5.7^{+1.5}_{-1.2}$ over
$[0.7, 1.2] R_{200}$.

These observational results are supported from the hydrodynamical simulations
of X-ray emitting galaxy clusters performed with the Tree+SPH code GADGET-2
(Roncarelli et al. 2006; see, e.g., Fig.~\ref{fig:simu}).
In the most massive systems, we measured a steepening of $S_b(r)$, independently
from the physics adopted to treat the baryonic component, with a slope of
$-4, -4.5, -5.2$ when estimated in the radial range $0.3-1.2 R_{200}$,
$0.7-1.2 R_{200}$, $1.2-2.7 R_{200}$, respectively.
In particular, we note the good agreement between the slope of the simulated
surface brightness profile of the representative massive cluster
in the radial bin $0.7-1.2 R_{200}$ (see values of $b_A$ in Table~4 of
Roncarelli et al. 2006 ranging between $-4.29$ and $-4.54$)
and the mean extrapolated value at $R_{200}$ of $-4.43$ measured in the
\chandra\ dataset.

\subsection{Temperature and metallicity profiles}\label{sec: tprof}

\begin{figure*}
\begin{center}
 \includegraphics[height=5.5cm]{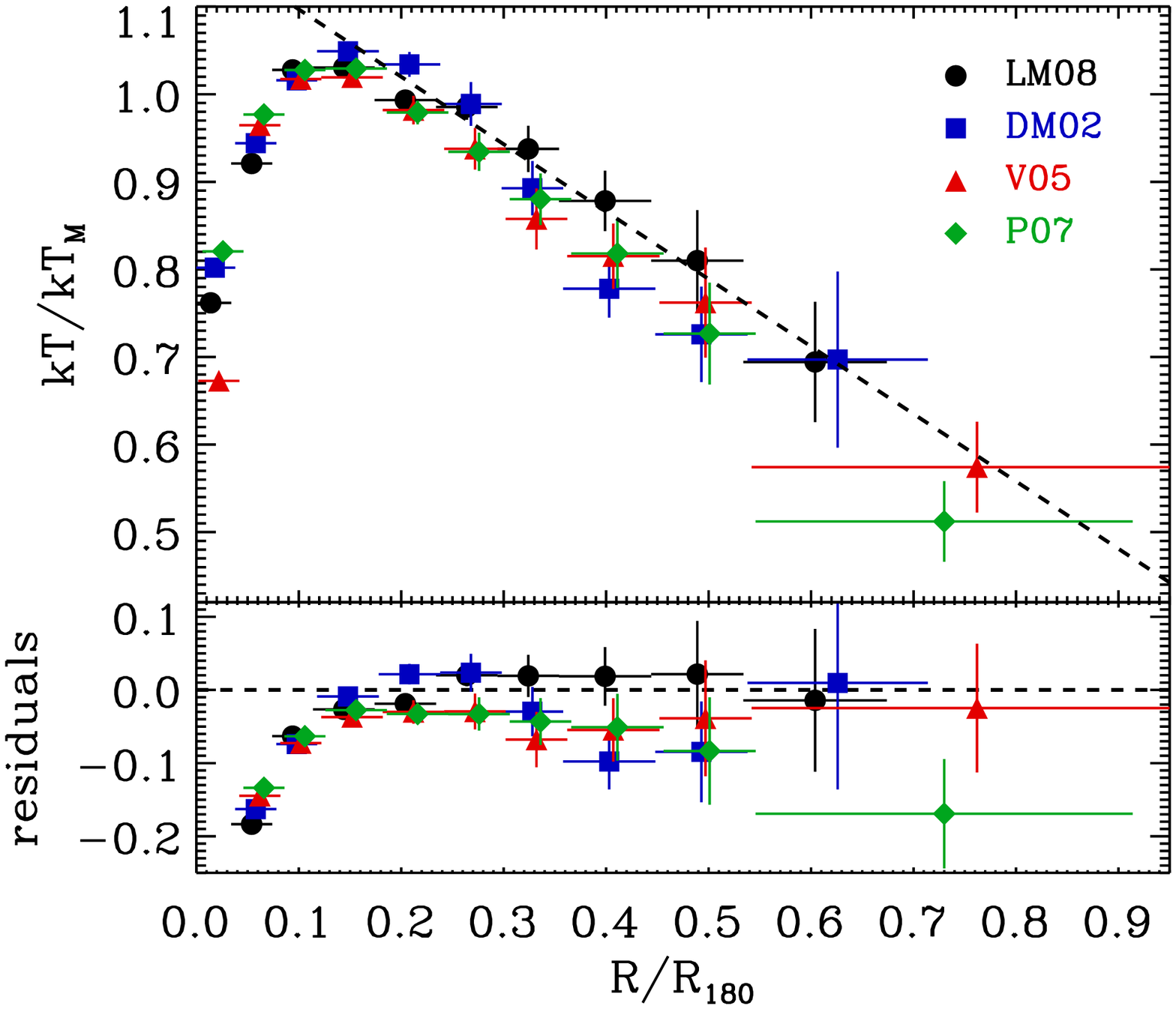}
 \includegraphics[height=5.7cm]{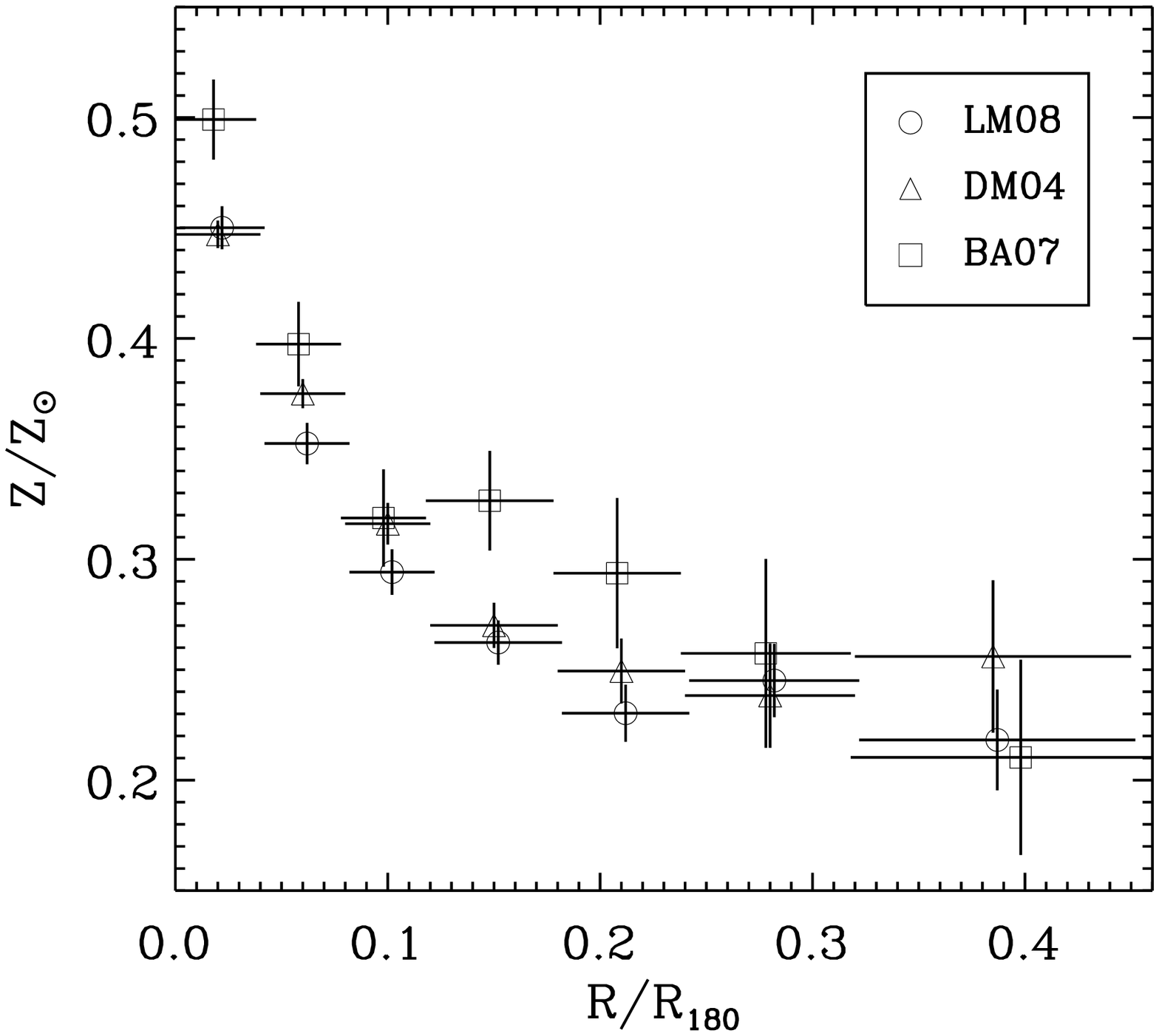}
\end{center}
\caption{\small  {\bf Left panel:} mean temperature profiles obtained from
  Leccardi \& Molendi (2008; LM08, black circles), De Grandi \& Molendi (2002;
  DM02, blue squares), Vikhlinin et al. (2005; V05, red upward triangles)
  and Pratt et al. (2007; P07, green diamonds).
All profiles are rescaled by
  $\mathrm{k}T_\mathrm{M}$ and $R_{180}$. The dashed line shows the best
  fit with a linear model beyond 0.2~$R_{180}$  and is drawn to guide the eye.
  {\it lower panel:} residuals with respect to the linear model.
  The LM08 profile is the flattest one.
  {\bf Right panel:} mean metallicity profiles obtained from Leccardi \& Molendi (2008b; LM08, circles),
  De Grandi et al. (2004; DM04, triangle) and Baldi et al. (2007; BA07, squares).
Abundances are expressed in Anders \& Grevesse (1989) solar
  values and radii in units of $R_{180}$. The radii have been slightly
  offset in the plot for clarity.
} \label{fig:prof_t}
\end{figure*}

Early attempts to produce temperature profiles  were made with the \rosat\ PSPC,
these were mostly limited to low mass systems (e.g. David et al. 1996) where the
temperatures were within reach of the PSPC soft response.
Resolved spectroscopy of hot systems began with the coming into operation of \asca\ (1994) 
and \sax\ (1996).
Both missions enjoyed a relatively low instrumental background, which was a considerable asset
when extending measures out to large radii, however they both suffered from limited spatially resolution.
The situation was somewhat less severe
with the \sax\ MECS than with the \asca\ GIS since the former had a factor of 2 better angular resolution
and a modest energy dependence in the PSF. These difficulties led to substantial differences in temperature
measures, while on the one side Markevitch et al. (1998) using \asca ~ and De Grandi \& Molendi (2002) 
using \sax\ MECS found evidence of declining temperature profiles, on the other,
White (2000) using \asca\ and Irwin et al. (1999) using \sax\ data found flat temperature profiles.
The situation was somewhat clearer on abundance profiles were workers using \asca\ 
(e.g. Finoguenov et al. 2000) and \sax\ data (De Grandi \& Molendi 2001) consistently found 
evidence that cool core systems featured more centrally peaked profiles than NCC system.
The coming into operation of the second generation of medium energy X-ray telescopes, 
namely \xmm\ and \chandra,
both characterized by substantially better spatial resolution, allowed more direct measures of the
temperature profiles.
The new \chandra\ (Vikhlinin et al. 2005) and \xmm\ measurements (e.g. Pratt et al. 2007, 
Snowden et al. 2008) confirmed the presence of the temperature gradients measured with \asca\ and \sax.
In a detailed study of a sample of 44 objects observed with \xmm\ (Leccardi \& Molendi 2008) we
found that temperature measurements could be extended out to about 0.7$R_{180}$ (see Fig.~\ref{fig:prof_t}).
Since the major obstacle to the extension of measurements to large radii was the high background, most importantly the
instrumental component, we adopted the source over background criterion originally introduced in De Grandi \&
Molendi (2002) to decide where to stop measuring profiles.
The source to background ratio, defined as ${I_{sou} \over I_{bkg}} $, where $I_{sou}$ and $I_{bkg}$
are the source and background intensities respectively, should not be confused with the signal to
noise ratio defined as  $ {I_{sou}\over (I_{sou} + I_{bkg})^{1/2}} \cdot t $, where $t$ is the exposure time.
While the latter ratio is associated to the statistical error and therefore increases with exposure time,
the former is associated to the systematic error and does not depend on the exposure time.
Through a series of tests
(see Sect.~5.2.1. and Fig.~11 of Leccardi \& Molendi 2008) we determined that measurements could be trusted
out to radii where the source to background ratio in the 0.7-10 keV band remained above a threshold of 0.6.
In  Leccardi \& Molendi (2008) we made use for the first time of extensive simulations to estimate the impact
of systematic errors on the measurements, part of the expertise we have acquired from that work has been
used to perform the simulations discussed in Sect.~\ref{sec: simul}.
In the left panel of Fig.~\ref{fig:prof_t} we show a compilation of mean temperature profiles from different
missions, all show evidence of a decline of the temperature beyond 0.2$R_{180}$. Interestingly,
as a result of the correction for systematic that we applied to our profile
(see Sect.~5.3 and Fig.~14 of Leccardi \& Molendi 2008)
ours is the flattest amongst the profiles shown in the left panel of Fig.~\ref{fig:prof_t}.
The measurement of the metal abundance profile extends to radii that are somewhat smaller than
those reached by the temperature profiles, this is because the most prominent emission line,
the Fe K$\alpha$, is located in the high energy part of the spectrum where the instrumental
background is particularly strong. In the right panel of Fig.~\ref{fig:prof_t} we show the mean abundance profile
measured with different satellites. The flattening of the profiles beyond
0.2$R_{180}$ is most likely indicative of an early enrichment of the ICM (Fabjan et  al. 2010).

Unfortunately the high orbit of the \xmm\ and \chandra\ satellites, as well as the fact that 
the design of the satellites was driven by scientific objectives other than the characterization 
of low surface brightness regions,
led to a substantially higher and more variable background than with the previous satellite generation,
thereby limiting the exploration of the temperature and abundance profiles to roughly the same regions
already investigated with \asca\ and \sax\ (see Fig.~\ref{fig:prof_t}).
Recently measures of temperature profiles have been made with the \suzaku\ X-ray imaging spectrometer (XIS).
Although not ideal for cluster measurements, the XIS features a poor PSF and a small FOV, it does enjoy the
considerable advantage of the modest background associated to the low earth orbit.
The measures have been conducted on a handful of systems (A2204, Reiprich et al. 2009;
A1795, Bautz et al. 2009; PKS0745-191, George et al. 2009; A1413, Hoshino et al. 2010)
and extend beyond the regions explored with \chandra\ and \xmm. However, the
characterization is a limited one at best: only  parts of the outermost annuli are explored and
both radial bins and error bars are large. Moreover there are concerns as to the reliability of the
measurements themselves.
All measured temperature profiles are steeper than those predicted by simulations.
This is particularly true of A1795 and PKS0745-191, where the temperature and the surface brightness
are respectively steeper and flatter than those predicted by simulations.
Consequently entropy profiles are flatter and, in the case of PKS0745-191, it features an inversion
around $0.6 R_{200}$, that could be associated to the presence of non virialized gas or, alternatively,
to problems in the characterization of the source spectrum.

\section{How we can map out to $R_{200}$}\label{sec: map}

\begin{figure*}
\begin{center}
 \hspace{2mm}\includegraphics[height=6.0cm,angle=-90]{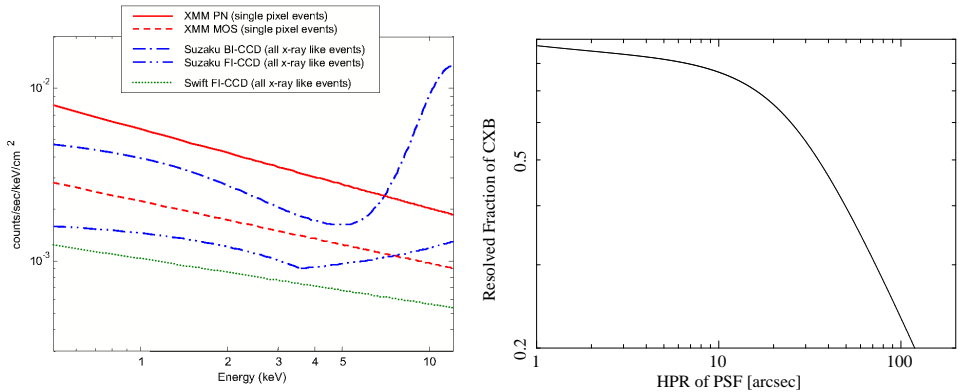}
 \hspace{2mm}\includegraphics[viewport = -50 0 553 662,height=6.9cm,angle=-90]{res_frac_ang2.ps}
\end{center}
\vspace{-11mm}
\caption{\small {\bf Left panel:} compilation of instrumental background spectra, only continuum components,
for various X-ray missions equipped with CCD detectors from Hall et al. (2008).
{\bf Right panel:} Resolved fraction of extragalactic cosmic X-ray background in
the 0.5--2 keV band as a function of angular
resolution.The total background intensity is derived from measures by De Luca \& Molendi (2004)
and Mc Cammon et al. (2002). 
The LogN-logS is taken  from Moretti et al. (2003). We also include an euclidian component to
account for  the unresolved 20\% of the cosmic X-ray background (CXB). The normalization of this component is
conservatively chosen in such a way that about 10\% of the CXB is found at fluxes that
are sufficiently small to efficiently mimic a diffuse component.
The angular resolution necessary to reach a given flux limit is obtained by imposing that
the source density at that flux limit is such that there is 1  source every 20 angular resolution
elements, where the angular resolution element is a circle with a radius equal to the half-power-radius
of the PSF.}\label{fig:bkg}
\end{figure*}

From the discussion in Sect.~\ref{sec: tprof}, it is rather obvious that past X-ray mission were
not optimized for the spectral characterization of the low surface brightness
emission typical of cluster outer regions.
In this section we discuss how to design
an experiment characterized by  high sensitivity to low surface brightness emission.
The sensitivity depends upon:
1) the surface brightness of the source, $S_{b}$, that scales with
effective area of the experiment, $A_{E}$;
2) the solid angle covered by the field of view (FOV), $\Omega$;
3) the surface brightness of the background, $B$.
The quantity that needs to be maximized is then:
$$ s = {2\pi\int_0^{\theta_{max}} A_{E}(\theta) \theta d\theta \over B}, $$
where $\theta$ is the off-axis angle and the integration is extended
over the full FOV, i.e. $2\pi\int_0^{\theta_{max}}\theta d\theta = \Omega$.
Therefore one needs to maximize the numerator,
$ 2\pi\int_0^{\theta_{max}} A_{E}(\theta) \theta d\theta $, a quantity that is
often referred to as ``grasp'',  and minimize the background
\footnote{A substantial fraction of the background is of instrumental
origin. This part and $A_{E}$ scale with the square of the focal length.
Both of them appears in the quantity $s$ that has to be maximized}.
To go well beyond what has been achieved with
the instrumentation that has been designed thus far one needs to
operate at three different levels: 1) the experiment design; 2) the observational
strategy; 3) the data analysis strategy.

\subsection{Experiment design}

Let us start by considering the background and in particular the instrumental background,
i.e. the part of the background that is not associated to genuine cosmic X-ray photons.
A few things can be easily inferred by comparing background spectra from different mission.
In Fig.~\ref{fig:bkg} we report a recent compilation of such spectra from Hall et al. (2008).
We note that: 1) front illuminated CCDs have lower background than background illuminated
ones and that  2) the background on the low earth orbit is smaller than that in the high
orbit. In this respect it is particularly instructive to compare the EPIC MOS with the \swift\
XRT background, since we are dealing with virtually the same detector in a high and low
earth orbit. As shown in Hall et al. (2008), the \swift\ XRT background is about a factor 3 lower
than the EPIC MOS background.
Thus, from the inspection of  Fig.~\ref{fig:bkg} we learn  that to keep the instrumental background low
it is preferable to employ front illuminated CCDs on a low earth orbit.
There are other issues that should be kept in mind: 1) a non-negligible fraction
(say 15\%) of the detector should be shielded from the sky, this will allow to
constantly monitor the intensity of the instrumental background;
2) a tilted CCD configuration which allows to improve the imaging, will result in fluorescence
Si line emission inhomogeneous distributed on the FOV, something similar is observed on
MOS EPIC, this can be minimized by studying the most appropriate configuration;
3) while active shielding cannot be applied as long as the detector is a CCD,
passive shielding can and should be considered. Most importantly the whole instrumental
background issue should be addressed from a global point of view. Detailed simulations of
the physical interaction between particles and photons with the satellite, possibly complemented by
exposures of the detector and associated structures to real particles and high energy photons, can be used
to study solutions that will minimize the background.

\begin{figure*}
\begin{center}
 \includegraphics[viewport = 91 0 555 662,clip,height=10cm,angle=-90]{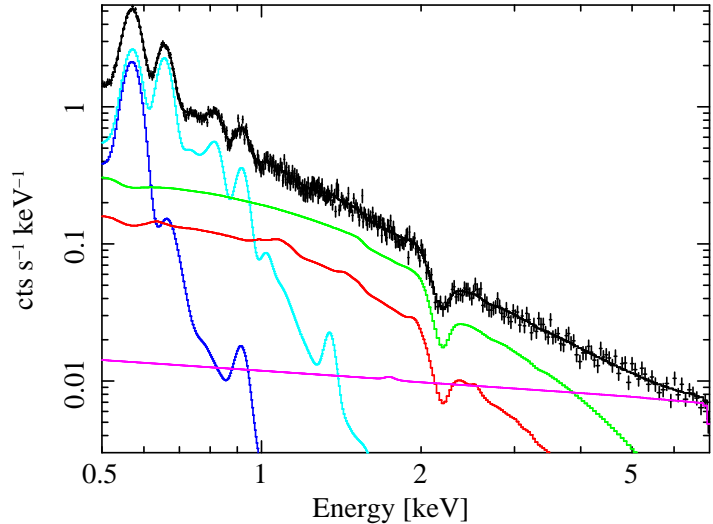}
\end{center}
\caption{\small Simulation of source and background spectra for a typical region at R200.
Tne instrumental background estimates come from current missions.
The cosmic X-ray background (CXB) is modeled with 3 components,
for the soft X-ray background (SXRB) we adopt the modeling of the SXRB
from McCammon et al. (2002), who
have carried out the highest spatial resolution observation of the
SXRB with a sounding rocket flight. The SXRB is modeled  by 2 thermal components
with temperatures of 0.1 keV and 0.225 keV  both with solar
abundances, normalizations  come from Table 3 of McCammon et al. (2002).
For the extragalactic background, comprising mostly unresolved AGNs,
we assume a power-law of slope
1.4 and intensity 1/4 of that derived by De Luca \& Molendi (2004),
thereby assuming that 3/4 of the sources will be resolved out.
The source surface brightness is assumed to be
$3\times 10^{-16} $ erg cm$^{-2}$s$^{-1}$arcmin$^{-2}$, a typical value for
cluster outskirts (see Sect.~\ref{sec: sb} for a detailed discussion), the extraction region is
100 arcmin$^2$, a value of $2\times 10^{20}$cm$^{-2}$ is assumed for the equivalent
hydrogen column density, the cluster emission is modeled by a thermal
plasma with kT$=$3 keV and metal abundance Z$=$0.15 Z$_\odot$, the exposure time
was set to 100 ks. The model spectrum was convolved with response files
(effective are and redistribution matrix) provided by the \wfxt\ team.
} \label{fig:spec_tot}
\end{figure*}

If the experiment is properly designed then the instrumental background will
be low and the cosmic background important. Above $\simeq$ 1 keV the dominant
contributor to the cosmic X-ray background is the extragalactic background associated to
unresolved sources, mostly AGN.
Sufficiently high spatial resolution allows to resolve  out a sizeable fraction of
the sources producing the X-ray background (see Fig.~\ref{fig:bkg}b).
With a resolution of 5 arcsec (Half-Power-Ratio, HPR)
it is possible to resolve out about 80\% of the background, provided of course sufficient
counts are available to detect the sources. It should be noted that
beyond an angular resolution of 15 arcsec the resolved fraction is not very sensitive
to the resolution, see Fig.~\ref{fig:bkg}b.
Another important point is that, to fully exploit the advantage of a large field
of view, it is necessary that the high spatial resolution be available over
the full FOV, polynomial optics (Burrows et al. 1992) can provide this important feature.
Another important contributor to the background  is the so called straylight,
this is associated to X-ray photons from outside the field of view
which end up in the focal plane after reflecting only once
on the mirrors. The effect of straylight can be significantly mitigated
by introducing a pre-collimator in front of the telescope as was done
in the case of the \xmm\ optics.

\subsection{Observing and data analysis strategies}

An experiment design like the one described above contributes significantly in
improving the sensitivity to low surface brightness emission, however further steps
need to be taken to reach cluster outer regions.
This is quite apparent when looking at the spectral simulation reported in
Fig.~\ref{fig:spec_tot} (for details see the figure caption).
As can be seen background components of one kind or another dominate the spectrum
at all energies. In the 1-3 keV range the source intensity is about 1/3 of the total,
below 1 keV the galactic foreground dominates, while above 3 keV the residual
extragalactic and the instrumental background do.
These are of course estimates, for real clusters things may be a little different,
however we will inevitably have a background that outshines the source.
These are atypical conditions with respect to previous X-ray imaging missions.
To make reliable measures will require devising specific observing and analysis
strategies.
Clearly the strongest requirement is that the background be characterized as well
as possible, ideally one would like to measure the background associated to
the source without the source, which is of course impossible.
Considering that the instrumental component varies with time and that
the galactic foreground varies with position on the sky, it is important to
observe  the background almost at the same time and almost at the  same location
of the source. A similar strategy has been adopted, albeit for reasons different from
the ones considered here, by the \swift\ XRT experiment. During each 1.5 hour orbit,
\swift\ observes a source field and 3 or 4 background fields. Thus background fields
are observed almost simultaneously with the source field and with the same
instrument set up. Moretti et al. (2010) have shown that under these conditions
the instrumental background can be characterized to the 3\% level. Conversely,
when background fields from different epochs are used, only a 10-15\% level is achieved.
The optimal solution that may be applied in a future mission, or on \swift\ for that matter, would be to
use as part of the background fields, sky regions close to the source and dark earth observations.
The former would allow to perform a spatial characterization of the galactic foreground, while
the latter would permit a clean measurement of the instrumental background.
Observations of both source and background fields need to be conducted to a high precision.
Relative systematic errors on the spectra need to be kept at the few percent level.
This is not a trivial requirement to meet, particularly since at this level of precision
each detector element has to be considered as an independent detector.
Assuming that each detector element will be calibrated to a relative precision of
$\simeq 5\%$, systematics can be reduced to the desired level
by viewing each sky element with a large number of detector elements. Observing strategies
such as this have been used for decades in other bands of the electromagnetic spectrum when
the source signal is smaller than the background. As examples, one may consider ground based infrared
observations or cosmic microwave background measurements.

The comparison of the source plus background spectrum with the background spectrum is
typically done via subtraction. In recent years, workers concentrating on cluster outer
regions (e.g. Snowden et al. 2008, Leccardi \& Molendi 2008) are finding that modeling
is more effective. This is readily understood if one considers that the background
is made of different components each capable of varying independently of the others.
Unless there are good reasons to believe that the particular combination
of background components associated to the source and background fields
are next to identical, it is preferable to model the different components
allowing for variations in relative intensity.
Another issue that should be considered is that, under the atypical
conditions of cluster outer regions, the standard maximum likelihood
estimators commonly employed to derive physical parameters such
as emission measure and temperature do not always work properly.
In a recent paper
(Leccardi \& Molendi 2007), we have shown that the presence of a
significant background component can lead to a substantially biased
measure of the temperature.
In the same paper, we describe a few quick fixes.
Unfortunately, a  general solution, based on a more powerful statistical
estimator, has yet to be found.

\subsection{A budget for systematics}

Assuming that the above guidelines are followed, we expect to be able to maintain systematic
errors to within a few percent. In the following, we provide a breakdown of the expected errors.
A constant monitoring of the instrumental background by using the part of the detector
not exposed to the sky plus dark earth and background field observations entwined with
source observations should allow us to constrain this component to about 1\%
(as extrapolated from the results obtained on \swift\ XRT in Moretti et al. 2010).
The extragalactic component of the cosmic background is a residual component,
comprising unresolved sources and possibly a diffuse component.
For a typical flux limit of 10$^{-16}$ erg cm$^{-2}$s$^{-1}$ in the 0.5-2.0 keV band,
montecarlo simulations show that the cosmic variance for a 100 arcmin$^2$ field is less
than 1\% of the residual background component.
The galactic foreground will be monitored by performing observations of fields contiguous
to the source field. Moreover, observations over several 100 arcmin$^2$ should allow us
to characterize this component to about 3-5\%.
Finally, assuming a typical relative calibration accuracy of 5\% on individual detector
elements and the application of substantial dithering, we expect to reach an overall 
relative spectral calibration of about 1\%.

\subsection{Detailed predictions}\label{sec: simul}

\begin{table*}[htb]
\caption{\small
Relative errors (in percentage) and deviations $\epsilon$ from the input values
at 90\% confidence level on the parameters of interest (normalization $K$, plasma temperature $T$ and
metal abundance $Z$ of an {\tt apec} component in {\sl XSPEC} --Arnaud 1996)
after joint-fit analysis of spectra simulated with an exposure time of 50 ksec.
All the relative errors can be rescaled to different exposure times as
$\sim \sqrt{t_{\rm exp}}$.
$\beta_{20}$ indicates a $\beta$ value increased by 20 per cent.
CC (nCC) indicates a (no) Cooling-Cores Cluster.
The fluxes $f$ are in units of $10^{-12}$ erg/s/cm$^2$ in the band
$(0.1-2.4)$ keV and are collected from
http://bax.ast.obs-mip.fr/.
}
\hspace*{0.1cm}
\begin{tabular}{cccccccc}
\hline \\
inputs & & $K$ & $T$  & & $K$ & $T$ & $Z$ \\
 & & \multicolumn{2}{c}{fixed $Z$} & & \\
\hline \\
\multicolumn{8}{c}{Perseus (TURBOLENT/CC;
 $z=0.0178, f=1137.3, T=6.3 {\rm keV}, n_{\rm H}=1.5e21; R_{200}=1.9 {\rm Mpc}=88.2'$)} \\
$T=3.16, Z=0.15$  &&  8 $(+0.2\epsilon)$ & 15 $(+0.5\epsilon)$ &&
   16 $(+0.3\epsilon)$  & 17 $(-1.3\epsilon)$  &  $>$100 \\
$T=2, Z=0.15$ &&  8 $(-1.0\epsilon)$ & 8 $(+0.7\epsilon)$  &&
   17 $(-0.1\epsilon)$  & 13 $(-0.4\epsilon)$  & 44 $(-0.2\epsilon)$ \\
$\beta_{20}, T=3.16, Z=0.15$ &&  27 $(+1.1\epsilon)$ &  38 $(-0.7\epsilon)$ &&
   66 $(+0.4\epsilon)$ & 55 $(-1.4\epsilon)$   & $>$100 \\
$\beta_{20}, T=2, Z=0.15$ && 25 $(+1.3\epsilon)$ & 30 $(+0.3\epsilon)$ &&
   51 $(+0.9\epsilon)$ & 47 $(-0.2\epsilon)$  &  5 $(-1.0\epsilon)$ \\
\\
\multicolumn{8}{c}{A1689 (MERGING/nCC;
 $z=0.1810, f=14.5, T=10.1 {\rm keV}, n_{\rm H}=1.8e20; R_{200}=2.2 {\rm Mpc}=12.2'$)}\\
$T=5.05, Z=0.15$ && 6 $(-0.7\epsilon)$ & 23 $(+0.5\epsilon)$ &&
   20 $(-0.9\epsilon)$ & 27 $(+1.1\epsilon)$ & $>$100 \\
$T=2, Z=0.15$ && 6 $(+0.1\epsilon)$ & 7 $(<0.1\epsilon)$ &&
   13 $(+0.2\epsilon)$ & 6 $(-3.2\epsilon)$  & 35 $(-1.3\epsilon)$ \\
$\beta_{20}, T=5.05, Z=0.15$ && $>$100 & $>$100   &&   $>$100 & $>$100 & $>$100 \\
$\beta_{20}, T=2, Z=0.15$ && 55 $(+0.2\epsilon)$ & 14 $(-3.7\epsilon)$ &&
  $>$100 & $>$100 & $>$100 \\
\\
\hline
\label{tab:feasi}
\end{tabular}
\end{table*}

Our goal is to resolve the physical properties of the ICM in the virial regions
making proper use of the \wfxt\ (FOV with $R_{\rm WF} \approx 30'$).

Our strategy is to define a set of observations with reasonable exposure time ($\leq$ 50 ksec)
that can allow the study of the virial regions through
the spatial and spectral analysis with \wfxt.

First, we select objects with known X-ray properties (flux, temperature,
dynamical status) that can be good candidates for a single \wfxt\ exposure,
i.e. with an expected $R_{200} < R_{\rm WF} = 30'$.
We can also relax a bit this assumption requiring however that a given
exposure minimizes the risks in term of
(i) problems of intercalibration with other X-ray observatories for
measurements in known X-ray emitting regions, (ii) weak constraints
on the X-ray properties at $R_{200}$ due to the effect of
unexpected large scale structures.

We estimate $R_{200}$ from a given spectroscopic measurement of the gas
temperature by using the best-fit results in Arnaud et al.
(2005, cf. Table~2; similar results in Vikhlinin et al. 2006):
\begin{equation}
R_{200} = 1714 \times (T_{\rm gas} / {\rm 5 keV})^{0.5} \, E_z^{-1} h_{70}^{-1} \, {\rm kpc}
\label{eq:r200}
\end{equation}
with $E_z = \left[ \Omega_{\rm m} (1+z)^3 +\Omega_{\Lambda} \right]^{0.5}$ and
$\Omega_{\rm m}=1-\Omega_{\Lambda}=0.3$.
\begin{figure}
\begin{center}
\vspace{-4mm}
 \includegraphics[bb = 86 28 600 594,clip,height=8cm]{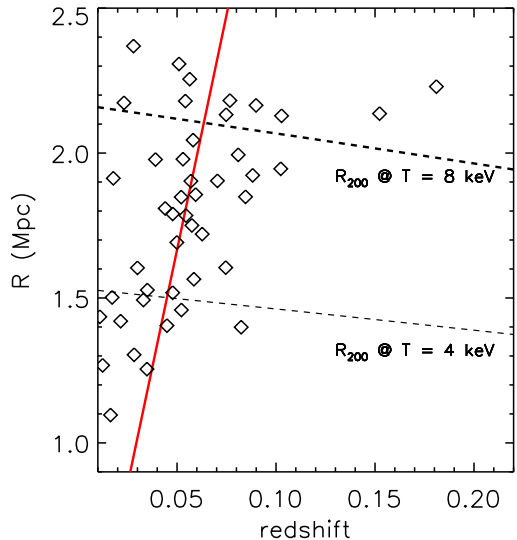}
\end{center}
\vspace{-4mm}
\caption{\small
Predicted $R_{200}$ for the 45 objects in Mohr et al. (1999) compared
to $R_{\rm WF} = 30'$ (red solid line) and the expected $R_{200}$
for a typical cluster with a temperature of 4 and 8 keV
(dashed lines). $R_{200}$ are estimated from eq.~\ref{eq:r200}.
} \label{fig:r200}
\end{figure}

By applying the criterion $R_{200} < R_{\rm WF} = 30'$,
we select 23 out of the 45 objects present in the flux-limited sample of
the brightest clusters in Mohr et al. (1999; see Fig.~\ref{fig:r200})).
More objects can be included if off-axis exposures are considered, as
requested for the Perseus cluster with a $R_{200}$ of $\sim 88$ arcmin.

The response matrix used for our simulations is obtained by convolving
the redistribution matrix with a mean effective area, $\overline{A(E)}$, constructed
by averaging the vignetting over the whole field of view, i.e.
$$ \overline{A(E)} = A_o(E) \cdot {2\pi\int_0^{\theta_{max}}\theta~ d\theta ~V(E,\theta)
\over \pi \theta_{max}^2}, $$
where $A_o(E)$ is the energy dependent on-axis effective area, $\theta_{max}$ is maximum
off-axis angle and $V(E,\theta)$ is the energy and off-axis dependent vignetting.
The redistribution matrix, the on-axis effective area and the vignetting were kindly provided
by the \wfxt\ team.
We use our own script with the response matrix
to simulate a {\it (source$+$background)} and a {\it (background only)} spectrum
including in the latter one
(i) a 1 per cent random fluctuation in absorbing $n_{\rm H}$ value and in the normalization
of the instrumental background;
(ii) a 5 per cent random fluctuation propagated to the normalization and temperature
values of the two local background component (one absorbed, the other not), to the
normalization and photon-index value of the CXB.
The photon-index is allowed to vary between $1.4$ and $1.6$.
We assume that 80 per cent of the CXB is resolved.

The spectra are integrated for 50 ksec over an area of 100 arcmin$^2$
and then jointly fitted in the range $0.3-6$ keV.

The surface brightness in the band $(0.5-2)$ keV are obtained from the best-fit values
in Table~2 of Mohr et al. (1999) by evaluating the model prediction at $R_{200}$
as estimated in equation~1.
A more conservative estimate of the surface brightness is obtained by increasing
the $\beta$ value by 20 per cent, faking an expected steepening of the
surface brightness profile in the cluster outskirts, as recent observations and
simulations suggest (see Section~2).
This correction reduces the predicted surface brightness by a factor of 7 on average.

All the simulated spectra assume a metallicity of $0.15 Z_{\odot}$ and a temperature
equal to 0.5 (see Roncarelli et al. 2006) times the quoted value in Table~1
of Mohr et al. (1999). We also consider the cases with metallicity equal
to $0.05 Z_{\odot}$ and temperature of about 0.25 times the quoted values
(i.e. between 1 and 2 keV).

Our simulated spectra (e.g. Fig.~\ref{fig:spec_tot}) show that
we can reach typical uncertainties (90\% level of confidence) of $\le$ 20\%
on the normalization $K$ and temperature $T$ of the thermal spectra
(see Tab.~\ref{tab:feasi}).
Reasonable constraints ($\sim$ 40\%) on the metallicity $Z$ can be obtained
in the case the surface brightness profile in the outskirts is still well
reproduced from the models fitted to {\it ROSAT} PSPC data.

A steepening of the surface brightness profiles, as expected from the work
discussed in Sect.~2 and modeled here by increasing the value
of the outer slope $\beta$ by 20 \%, reduces significantly the level of accuracy to which
we can constrain the physical parameters: about 60 per cent (relative error
at 90\% level of confidence) on $K$, 40 per cent on $T$, no constraints on $Z$.

\section{Future missions \& \wfxt}\label{sec: future}

In this section we provide an overview of missions under study or construction that
may provide important contributions to the characterization of cluster outer
regions. There are 3 such missions namely SRG, \xenia\ and \wfxt.
The eROSITA experiment (Predehl et~al. 2007) on board the Russian Spektrum Roentgen Gamma (SRG) 
satellite comprises 7 telescopes with a total on-axis effective area of 2000 cm$^2$, 
an on-axis angular resolution of 25 arcsec and will operate from an L2 orbit. 
\xenia\ (Hartmann et al. 2009) carries an X-ray imager and spectrometer that would both be useful
in characterizing cluster outskirts: the imager has an on-axis effective area of 600 cm$^2$ and an
on-axis angular resolution of 15 arcsec; the spectrometer has an unprecedented spectral resolution of
a few eV, an on-axis effective area of about 1000 cm$^2$ and an angular resolution that is limited 
by the pixel size of a few arcmin. 
\wfxt\ (Murray et~al. 2010) which, like \xenia, has been submitted to the 
{\it Astro2010: The Astronomy and Astrophysics Decadal Survey}, carries an
X-ray imager comprising 3 telescopes for a total on-axis effective area of 6000 cm$^2$,
and an on-axis angular resolution of 5 arcsec (requirement for the half-power-radius
of the PSF). 
A low earth equatorial orbit is forseen for both \xenia\ and \wfxt.

Both the \xenia\ and \wfxt\ imager have two considerable advantages
over eROSITA, namely the low earth over the L2 orbit and the
polynomial optics, which will result in a substantial reduction of
the instrumental and cosmic X-ray background, respectively.
In particular, the \wfxt\ imager will provide the characterization of
the cluster outer regions in about 1/10 of the time requested from
\xenia, and will benefit from higher angular resolution. 
\xenia\, however, is in the unique position to complement the imager 
data with high spectral resolution data for relatively bright clusters.
While eROSITA is scheduled for launch in 2012, \xenia\ and \wfxt\ are
both at an early stage of development and have to be considered as
the next generation satellites for clusters studies.

\section{Summary}\label{sec: summary}
Past and current X-ray mission allow us to observe only a fraction of the volume
occupied by the ICM. Indeed, typical measures of the surface brightness, temperature and
metal abundance extend out to a fraction of the virial radius. The coming into operation of the
second generation of medium energy X-ray telescopes at the turn of the millennium,
has resulted in relatively modest improvements in our ability to characterize cluster
outskirts. Even though recent results from \suzaku\ show some improvement,
the  most sensitive instrument to low surface brightness to have flown thus far
is quite possibly the \swift\ XRT which, ironically, never had cluster outer regions
as one of its top scientific objectives.

The construction of an experiment capable of making measures out to $R_{200}$ is well
within the reach of currently available technology. What is required is an
experiment design that will minimize the background, both instrumental and cosmic,
and maximizes the grasp, i.e. the product of effective area and FOV.
Since cluster emission in the outskirts will be background dominated,
instrument design and observational strategy should also
allow for a meticulous characterization of the background.
Detailed simulations based on realistic estimates of the different spectral
components and of the precision with which the may be determined shows that
an experiments such as the one we envisage will allow a solid characterization
of cluster outskirts.
From what we can surmise \wfxt\ is already designed to meet most of the
requirements which are necessary to characterize cluster outskirts, and should have
no major difficulty in accommodating the remaining few.

\section*{ACKNOWLEDGEMENTS}
We acknowledge the financial contribution from contracts ASI-INAF
I/023/05/0 and I/088/06/0.

\end{document}